\documentclass[aps,pra,twocolumn,superscriptaddress,nofootinbib]{revtex4-1}

\usepackage{amsmath,amsthm,amsfonts,amssymb,bm}
\usepackage{epsfig,epstopdf,graphicx,nomencl,upgreek,url}
\usepackage[colorlinks={true},pdftex]{hyperref}  % use with PDF
\usepackage{color}
\hypersetup{
  colorlinks,
  citecolor=blue,
  linkcolor=blue,
  urlcolor=blue}

\newcommand{\beq}{\begin{equation}}
\newcommand{\eeq}{\end{equation}}
\newcommand{\bqa}{\begin{eqnarray}}
\newcommand{\eqa}{\end{eqnarray}}
\newcommand{\bqan}{\begin{eqnarray*}}
\newcommand{\eqan}{\end{eqnarray*}}
\newcommand{\nn}{\nonumber}

\newcommand{\erf}[1]{Eq.~(\ref{#1})}

\newcommand{\eg}{\emph{e.g.},~}
\newcommand{\ie}{\emph{i.e.},~}
\newcommand{\etal}{\emph{et al.}}

\newcommand{\dg}{^{\dagger}}

\newcommand{\SNLCA}{Sandia National Laboratories, Livermore, CA 94550, USA}
\newcommand{\SNLNM}{Sandia National Laboratories, Albuquerque, NM 87123, USA}
\newcommand{\Stanford}{Edward L. Ginzton Laboratory, Stanford University, Stanford, CA 94305, USA}

\allowdisplaybreaks[1]
\begin{document}

\title{Silicon nanophotonics for scalable quantum coherent feedback networks}
\author{Mohan Sarovar}\email{mnsarov@sandia.gov}\affiliation{\SNLCA}
\author{Daniel B. S. Soh}\affiliation{\SNLCA}\affiliation{\Stanford}
\author{Jonathan Cox}\affiliation{\SNLNM}
\author{Constantin Brif}\affiliation{\SNLCA}
\author{Christopher T. DeRose}\affiliation{\SNLNM}
\author{Ryan Camacho}\affiliation{\SNLNM}
\author{Paul Davids}\affiliation{\SNLNM}

\begin{abstract}
The emergence of coherent quantum feedback control (CQFC) as a new paradigm for precise manipulation of dynamics of complex quantum systems has led to the development of efficient theoretical modeling and simulation tools and opened avenues for new practical implementations. This work explores the applicability of the integrated silicon photonics platform for implementing scalable CQFC networks. If proven successful, on-chip implementations of these networks would provide scalable and efficient nanophotonic components for autonomous quantum information processing devices and ultra-low-power optical processing systems at telecommunications wavelengths. We analyze the strengths of the silicon photonics platform for CQFC applications and identify the key challenges to both the theoretical formalism and experimental implementations. In particular, we determine specific extensions to the theoretical CQFC framework (which was originally developed with bulk-optics implementations in mind), required to make it fully applicable to modeling of linear and nonlinear integrated optics networks. We also report the results of a preliminary experiment that studied the performance of an \emph{in situ} controllable silicon nanophotonic network of two coupled cavities and analyze the properties of this device using the CQFC formalism.
\end{abstract}
 
\maketitle

\section{Introduction}

Over the last decade, coherent quantum feedback control (CQFC) has emerged as a new interdisciplinary field in the areas of quantum control and quantum engineering, and enjoyed rapid theoretical and experimental advances. In particular, a powerful theoretical framework based on input-output theory has been developed for modeling networks of quantum systems connected by electromagnetic fields \cite{Gar.Col-1985, Gar-1993, Wis.Mil-1994a, Zhang:2012jr, Gough:2012fl, CombesKerckhoffSarovar2016}. Such networks can be designed to operate as autonomous devices for quantum information tasks, e.g., quantum state preparation and stabilization \cite{Kerckhoff:2010cn, Crisafulli:2013dq, Hamerly:2013kb}, as well as ultra-low-power optical processing elements for applications such as optical switching \cite{Mabuchi:2009ck, Santori:2014tr}. Recent developments such as the SLH formalism for modular analysis of optical networks \cite{Gou.Jam-2009, Gou.Jam-2007} and the QHDL language and QNET software tools for specification and simulation of photonic circuits \cite{Tezak:2012uc} have added important capabilities for efficient and automated design and modeling of CQFC networks. 

These advances bring CQFC to the level where it can help design and quantitatively analyze quantum effects in integrated optics systems and, reciprocally, benefit from fabrication capabilities of  nanophotonics technology. Indeed, the true potential of CQFC is realized at scales where many optical elements are interconnected as potentially reconfigurable networks~\cite{Kerckhoff:2011vf}. Bulk-optics implementations are impractical for the realization of such complex networks, and a form of integrated optics is necessary. In addition to the obvious size advantage of integrated optics, other benefits include reproducibility and mass production capability, and long-term optical path and phase stability. In particular, CMOS-compatible silicon integrated nanophotonics is seen as a leading platform for constructing large-scale CQFC networks.

In this paper, we present a detailed analysis of the potential of silicon nanophotonics for implementing CQFC networks. We describe the behavior of nanophotonic components relevant for construction of CQFC networks and also discuss the key challenges to silicon nanophotonics implementation of CQFC. Finally, we describe our fabrication and measurement of a simple on-chip CQFC network composed of two coupled cavities and analyze it using the SLH formalism. As many of the general points raised in the paper are revealed in the analysis of this device, this implementation serves as a useful testbed on which to explore the benefits and challenges of silicon nanophotonics realizations of CQFC.

The remainder of the paper is structured as follows. Section~\ref{sec:slh} presents a brief introduction to the theoretical framework that enables efficient modeling of optical networks composed of modular components connected by quantum fields. In Sec.~\ref{sec:onchip} we discuss the current capabilities of CMOS-compatible integrated optics platforms, including the common linear and nonlinear components that are available for constructing quantum optical networks. Section~\ref{sec:theory} explores potential challenges in extending the standard CQFC theory presented in Sec.~\ref{sec:slh} to the integrated photonics realm. In Sec.~\ref{sec:device}, we present a preliminary on-chip implementation of a network of two coupled cavities, and analyze its properties. Finally, Sec.~\ref{sec:disc} concludes with a discussion of potential future directions in integrated optics CQFC.

\section{The SLH formalism for modeling CQFC networks}
\label{sec:slh}

In this section, we provide a brief summary of the approach to modeling CQFC networks using quantum stochastic calculus, or alternatively input-output theory from quantum optics (for further details, see Refs.~\cite{Zhang:2012jr, Gough:2012fl, CombesKerckhoffSarovar2016}). The basis of this approach is the decomposition of a quantum optical network into localized components with arbitrary internal degrees of freedom, which are connected via freely propagating unidirectional broadband fields. This allows one to eliminate the explicit description of the fields propagating between components to arrive at an effective description of the network just in terms of the localized degrees of freedom and their couplings. We refer to this modular approach and the associated modeling machinery as the SLH formalism~\cite{Gou.Jam-2009, Gou.Jam-2007}.

The starting point for this formalism is the Hudson-Parthasarathy theory that uses a quantum stochastic differential equation (QSDE) to represent time evolution of the unitary operator, $U(t)$, describing coupled evolution of the system and field degrees of freedom \cite{Hudson:1984}:
\bqa
dU(t) &=& \left\{ (S-1)d\Lambda(t) + L ~dB\dg(t) - L\dg S dB(t) \right. \nn \\
&& \left. - \left(\tfrac{1}{2} L\dg L + i H \right) dt \right\} U(t). 
\label{eq:hp_one}
\eqa
Here, $B(t)$ and $B\dg(t)$ are integrated versions of the freely propagating bosonic fields linearly interacting with the system at an interface or ``port" (these could be output fields from another system):
\bqa
B(t) = \int_0^t b(s) ds, ~~~~ B\dg(t) = \int_0^t b\dg(s) ds,
\eqa
with $[b(t), b\dg(s)] = \delta(t-s)$. This commutation relation defines the bosonic fields as rather singular objects, and hence the increments, $dB(t)=B(t+dt)-B(t)$ (and similarly $dB\dg(t)$), are operator-valued stochastic variables that are analogous to Ito increments. Finally, $\Lambda(t)$ is a quantum stochastic process that corresponds to the observable counting the number of quanta in the bosonic field that have interacted with the system up to time $t$:
\beq
\Lambda(t) = \int_0^t b\dg(s) b(s) ds.
\eeq
The other components of \erf{eq:hp_one}, the system operators $S$, $L$, and $H$, describe the system and its interaction with the propagating field at the interface. Specifically, $S$ describes the impact on system when photons are scattered between ports (this component is most interesting when we consider systems with multiple ports, as we shall below), $L$ is the system operator that is directly and linearly coupled to the field, and $H$ is the system Hamiltonian that accounts for the internal dynamics that does not involve interaction with the field. These components are often grouped together into a triple $G = (S,L,H)$, which is sufficient to completely characterize the system evolution. 

The generalization of \erf{eq:hp_one} to the case where the system has multiple ports, with independent fields at each port interacting with system, is:
\bqa
dU(t) &=& \left\{ \sum_{jk} (S_{jk} - \delta_{jk}) d\Lambda_{jk}(t)  \right. \nn \\
&& + \sum_j L_j dB_j\dg(t) - \sum_{jk}L_j\dg S_{jk} dB_k(t) \nn \\
&& \left. - \left( \frac{1}{2}\sum_{j}L_j\dg L_j + i H \right) dt \right\} U(t), 
\eqa
where $S_{jk}$ describes the effect on the system of a photon scattering from port $j$ to $k$, and $L_j$ is the system operator coupled to the field at port $j$. In the multi-port case, we still describe the system evolution using an SLH triple, but now $S$ is an $n \times n$ matrix and $L$ is an $n \times 1$ matrix (where $n$ is the number of ports) containing operator-valued elements: 
\bqa
S = \left(\begin{array}{ccc}S_{11} & \dots & S_{1n} \\ \vdots & \ddots & \vdots \\S_{n1} & \dots & S_{nn}\end{array}\right),
\quad L = \left(\begin{array}{c}L_1 \\ \vdots \\ L_n \end{array}\right) .
\eqa

The key advantage of the SLH formalism is that one can easily construct effective descriptions of arbitrarily connected networks of localized components, each of which is represented by a triple: $G=(S,L,H)$. Connecting two components in series, parallel, or feedback loop results in another system represented by another SLH triple whose matrices can be derived by simple algebraic rules~\cite{Gou.Jam-2009}. As an example, consider connecting two localized systems in series, where the outputs from $G_1 = (S_1, L_1, H_1)$ are connected to the inputs of $G_2 = (S_2, L_2, H_2)$, where for simplicity we assume that the number of input ports that $G_2$ has is the same as the number of output ports that $G_1$ has. The resulting system is represented as $G_3 = G_2 \lhd G_1 = (S_3, L_3, H_3)$, where $S_3 = S_2 S_1$, $L_3 = S_2 L_1 + L_2$, and $H_3 = H_1 + H_2 + \textrm{Im}\left\{L_2\dg S_2 L_1 \right\}$. See Refs.~\cite{Gou.Jam-2009, Gough:2012fl} for more details on the composition rules for the SLH formalism.

Once the SLH triple for a network of components has been calculated, the output fields from the network are easily calculated through the prescription:
\beq
dB^{\rm out}_{k}(t) = L_k(t) dt + \sum_{l} S_{kl} dB^{\rm in}_l (t)
\eeq

Finally, since we are interested in applying the SLH formalism to integrated optics networks, it is useful to explicitly list the most relevant assumptions used in developing the formalism:
\begin{enumerate}
%\item The system and propagating fields interact linearly at each interface/port.
\item The propagating fields are bosonic (although extensions of the formalism to fermionic fields are possible \cite{Gardiner:2004en}).
\item The interaction between the system and a field --- and the system and any other reservoir --- is Markovian, in the sense that the strength of the interaction is independent of the frequency of the field mode, at least for a reasonably wide band of frequencies. See \cite[Sec.~5.3]{Gar.Zol-2004} for a formal specification of this assumption.
\item The SLH composition rules assume that fields propagate between localized system components in a lossless, dispersionless, linear medium, and furthermore that the propagation time is negligible compared to timescales relevant to the localized systems.
\end{enumerate}

We will discuss the validity of these assumptions for integrated optics in the following sections.

\section{On-chip optical elements for CQFC networks}
\label{sec:onchip}

Solid-state based realizations of quantum optical networks are possible on several platforms nowadays, including microwave quantum optics on superconducting integrated structures \cite{Wal.Sch.etal-2004, Kerckhoff:2013iy, Liu:2016kb} and visible/near infrared quantum optics on integrated photonic structures \cite{OBrien:2009eu}. In this work, we focus on the latter and, in particular, on integrated photonics implementations using silicon and silicon nitride at telecommunications wavelengths centered on 1550~nm. The CMOS compatibility and relative maturity of integrated photonics on these platforms makes them appealing candidates for implementing scalable CQFC networks. See Combes \etal \cite{CombesKerckhoffSarovar2016} for a review of integrated implementations of quantum optical networks, and a discussion of the issues relevant to superconducting circuit implementations. 

Silicon (Si) and silicon nitride nanowire waveguides guide light through total internal reflection, enabled by the high index contrast between the guiding core and the surrounding cladding. Usually the waveguide is designed to guide only a fundamental single mode at a desired wavelength. The shape of the waveguide is fully etched, which naturally fits with CMOS compatible processes. Since the waveguide is rectangular with greater width than height, the profile of the guided mode of light is elliptical. The large index contrast between the core and the cladding allows for waveguide dimensions to be only a fraction of the wavelength (several hundreds of nanometers).  

A crucial factor that differentiates transmission in silicon waveguides from transmission in vacuum is scattering of photons due to roughness of waveguide surfaces. This leads to a linear loss mechanism (loss that is independent of light intensity) in waveguides and resonant structures. In the nanowire type single mode waveguides, scattering loss arises due to side-wall inhomogeneities created in the etch process used to define the waveguide. This scattering can be minimized by sophisticated fabrication techniques~\cite{Alasaarela:2011, Bojko:2011, Saynatjoki:2011, Wood:2014}, but is an intrinsic non-ideality that cannot be completely mitigated.  In the nitride nanowire waveguides, absorption due to unsaturated bonds is another source of propagation loss and can be mitigated by sophisticated passivation, deposition and fabrication techniques, \eg \cite{Borselli:2007ga,Mao:2008fh}.
   Other waveguide types,  such as ridge waveguides, can have lower scattering loss although they are intrinsically multi-mode.  These ridge modes are usually more tightly confined and hence interact weakly with side-wall roughness. All integrated optical waveguides and resonant structures will have intrinsic losses that cannot be completely mitigated. 

\subsection{Linear optics elements}
\label{sec:linear_opt}

Nearly all linear optical elements have been realized in silicon integrated optics, and in Table~\ref{tab:lin_optics} we list common linear bulk-optics elements and their integrated-optics counterparts. One element that requires particular attention in this list is the integrated optics cavity. These cavities are typically resonant structures such as microring resonators that result in high field intensity in a localized region. This high field intensity implies that nonlinear effects cannot be ignored. Effectively, this means that any integrated optics cavity with a sufficiently high quality factor ($Q$ factor) must be treated as a nonlinear element, rather than a simple linear cavity. We discuss in more detail the conditions for nonlinear dynamics in integrated resonant structures in the next section.

\begin{table*}[htbp]
\caption{\label{tab:lin_optics}Common linear bulk-optics elements and their integrated-optics counterparts.}
\begin{ruledtabular}
\begin{tabular}{|p{0.14\linewidth}|p{0.29\linewidth}|p{0.54\linewidth}|}
\textbf{Bulk optics} & \textbf{Integrated optics} & \textbf{Notes} \\ \hline
Beam splitter & Directional coupler & The transmissivity tuned by proximity of waveguides \\ \hline
Cavity & Microring resonator, whispering gallery mode resonator & The dimensions of integrated optics cavities are typically in the micrometers, allowing for very large (GHz) cavity bandwidths. Photon build-up and dissipation times are accordingly shortened, allowing for GHz switching. Furthermore, due to the reduced mode volume, integrated optics has great potential to demonstrate strongly coupled cavity quantum electrodynamics \cite{Lodahl:2015fy}. \\ \hline
Mirror & Distributed Bragg mirror & The reflectivity tuned through modulation depth and/or number of Bragg periods. \\ \hline
Phase shifter/modulator & Phase modulation by thermo-optic effect, or carrier injection/depletion & Carrier density manipulation achieves much greater frequency of phase modulation compared to the thermo-optic effect. \\ \hline
Amplitude modulator & Mach-Zehnder~Interferometer (MZI) through phase modulators & The amplitude of the MZI output field is controlled by varying the phase shift in one arm of the MZI. \\ 
\end{tabular}
\end{ruledtabular}
\end{table*}

\subsection{Nonlinear optics elements}
\label{sec:non-linear_opt}

Silicon and silicon nitride are highly nonlinear materials and a variety of optical nonlinearities with a typical Kerr coefficient $n_2 \sim 4 \times 10^{-18}$ m$^2$/W, a hundred times larger than the silica fibers, have been demonstrated on these platforms at 1550~nm \cite{Foster:2006}. Due to the centrosymmetry of these materials, $\chi^{(2)}$ processes are negligible and $\chi^{(3)}$ processes are the dominant sources of nonlinearity~\footnote{$\chi^{(2)}$ processes can be induced by introducing strain or interfaces to other materials, e.g., see Ref.~\cite{Levy:2011gs}.}. The intrinsic nonlinearity of silicon can be effectively enhanced by using structures that provide high optical field confinement, such as ring resonators. 

To get an idea for conditions where the nonlinearity of the guiding material must be taken into account, consider the nonlinear phase shift acquired by a light mode after traveling a length $\ell$ along a waveguide: $\Delta \phi_{\rm nl} = \gamma \ell P$, where $P$ is the peak power for a pulse and average power for a continuous-wave (CW) mode, and $\gamma = (2\pi n_2)/(\lambda A_{\mathrm{eff}})$ is the nonlinearity parameter defined in terms of the mode wavelength, $\lambda$, nonlinear refractive index of the material, $n_2$, and the effective area of the mode, $A_{\mathrm{eff}}$. If $\Delta \phi_{\rm nl} \gtrsim 0.1$, this nonlinear phase shift, and other nonlinear effects associated with propagation in the material such as wave mixing, cannot be ignored \footnote{This threshold of 0.1 is somewhat arbitrary, but its exact value does not significantly change the following argument.}. Assuming an input power of $P = 1$~$\mu$W (typical for CQFC applications), and a nonlinear coefficient of $\gamma = 1.5 \times 10^5$~W$^{-1}$~km$^{-1}$ for a strip waveguide in silicon guiding light at 1550~nm~\cite{Eggleton:2011gf}, we find that the nonlinearity is significant for waveguides of length $\ell \gtrsim 670$~m. This result shows that, under typical conditions relevant for CQFC applications, nonlinear effects are negligible for silicon integrated waveguides. However, in structures that localize and concentrate light, such as ring resonators, the circulating power is enhanced by a factor $B$ over the input power. In an ideal (lossless) resonator this enhancement factor is related to the resonator $Q$ factor by~\cite{Rabus:2007ul}
\beq
B = \frac{\lambda Q}{\pi n_{\mathrm{eff}} \ell_{\mathrm{r}} },
\eeq
where $\ell_{\mathrm{r}}$ is the circumference of the ring resonator. Using this expression for the power enhancement factor, an effective index of $n_{\rm eff}=2.85$, and an input power of $1$~$\mu$W, we find that the nonlinearity plays a role if the resonator $Q$ factor is $Q \gtrsim 3.5\times 10^9$. Therefore in very high quality resonant integrated structures one must be mindful of nonlinear effects. 

Even though significant progress has been made in utilizing nonlinear integrated optics elements for classical optical processing \cite{Leuthold:2010dg}, for example, lasing~\cite{Boyraz:2004, Rong:2005, Liang:2010}, parametric amplification~\cite{Foster:2006, Liu:2010, Kuyken:2011}, electro-optical modulation~\cite{Lipson:2005, Xu:2005, VanCamp:2012}, and frequency conversion~\cite{Hu:2011}, on-chip \emph{quantum} nonlinear optics is still in its infancy. The primary obstacle to realizing high quality nonlinearities for quantum optics on the silicon photonics platform is the need to overcome the deleterious processes that accompany advantageous nonlinear processes such as four-wave mixing (FWM). Two such deleterious processes that are particularly important in silicon are two-photon absorption (TPA) and dispersion. The first is a source of photon loss that can counteract any nonlinear gain \cite{Helt:2014wc}, while the second makes phase matching in quasi-one-dimensional waveguides challenging \cite{Leuthold:2010dg}. 

Despite these challenges, there have been several recent demonstrations of on-chip quantum nonlinear optics, for example, spontaneous FWM using a silicon microring resonator \cite{Azzini:2012kj}, and squeezed light generation using a silicon nitride microring resonator functioning as an optical parametric oscillator (through FWM) \cite{Dutt:2013wr}. We discuss the latter in more detail since the optical parametric oscillator (OPO) is an essential nonlinear component in many CQFC networks that have been proposed or constructed to date.

Dutt \etal~state~\cite{Dutt:2013wr} that silicon nitride was chosen for its lack of TPA and moderate nonlinearity. In addition to minimizing loss, they engineered several features in order to successfully exploit the FWM process in the material, including: (i) the resonator had very high intrinsic $Q$ factor while simultaneously being over-coupled to the output waveguide, which enabled field concentration as well as high bandwidth squeezing, (ii) the dispersion and quality factors of the resonator were engineered to yield wide spectrally spaced resonant modes, three of which could be selected as pump, signal, and idler modes (the wavelength separation enabled spectral isolation of the desired squeezed output modes), and (iii) a high numerical aperture optical lens was utilized to minimize the detection losses. As a result of these efforts, the detected signal and idler modes had intensity correlations 1.7~dB below the shot-noise level. 
The same group has recently also reported on a system of coupled silicon nitride cavities that were engineered to enable tuning of the degree of squeezing from 0.9~dB to 3.9~dB (on chip) \cite{Dutt:2016ky}.
Clearly there are more engineering challenges to on-chip squeezed light generation compared to its bulk-optics analogue, but given these demonstrations of experimental feasibility \cite{Dutt:2013wr,Dutt:2016ky}, we expect that material and design improvements will make OPOs a standard nonlinear integrated optics element in the near future.

\subsection{Sources and detectors}
Integrated photon sources within Si photonics are extremely limited.  This is due mainly to the fundamental issue of the direct band-gap of Si.  
Instead, the most widely adapted approach to generation of light on chip uses heterogenous integration of III-V laser gain materials bonded to Si as the laser cavity and/or transport media \cite{fang2006electrically}.  Recently, there have also been reports that strained and heavily $n$-type doped germanium (Ge) can be made to lase at telecom wavelengths \cite{liang2010recent}.  The effect of compressive strain on Ge is to split the light hole and  heavy hole bands effectively shrinking the direct band-gap of Ge.  The $n$-type doping is used to fill the $L$-valley indirect band-gap effectively Pauli blocking these states and in essence making the Ge a direct band-gap material.  These Ge laser devices are typically optically pumped and due to the highly doped nature of the Ge and the resulting free-carrier absorption require excessive power resulting in thermal destruction of the devices during operation.  Other integrated sources include germanium-tin (GeSn) \cite{wirths2015lasing} and highly strained Ge \cite{jain2012micromachining} which emit at longer infrared wavelengths outside of the telecom band and are currently the subject of much work on mid-infrared photonics.  

The state of integrated detectors is completely different from that of integrated sources.  Many groups have demonstrated integrated on-chip high performance Ge on Si photodiodes \cite{Aminian2012,Derose2011,Assefa2010,Bowers2010,Kang2008}.  Standard  Ge on Si p-i-n photodiodes and  separate absorption charge multiplication linear mode avalanche photodiodes have been integrated into Si photonics processes and demonstrated  record performance \cite{Derose2011,martinez2016high}. Geiger mode operation of these top-illuminated devices have also shown single photon detection with efficiencies ranging from 5 to 10\% \cite{6620943,hadfield2009single}.  Furthermore, superconducting nanowire single photon detectors (SNSPDs) based on W silicide \cite{marsili2013detecting} and niobium and niobium nitride superconducting films \cite{natarajan2012superconducting} are completely compatible with advanced Si photonics  manufacturing processes and  have demonstrated waveguide coupled performance in excess of 93\% \cite{pernice2012high}.  It therefore remains to develop an integrated process to incorporate SNSPDs with cryogenically compatible Si photonics.

\subsection{Isolators}
A standard bulk optics element that is \emph{not} commonly available in silicon integrated photonics is an isolator. Optical isolators minimize back reflection along a channel and are critical for defining unidirectional fields, which is especially important in feedback loops where a clear direction of signal flow is required. Despite some recent progress in constructing an on-chip optical isolator \cite{Bi:2011ef, Sounas:2013ey, Jalas:2013hk}, this remains a difficult element to incorporate into the integrated photonics toolbox, and other techniques for ensuring unidirectional propagation are necessary. For example, back reflection from cavity interfaces can be minimized by side coupling waveguides to ring resonator cavities with minimal surface roughness. Also, for a train of optical pulses, it is possible to reduce back-scattering using a timed add-drop multiplexer (although this requires that the arrival window of the back-scattered or reflected light pulse is known) \cite{Agarwal:2010td}. 

\section{Application of the SLH formalism to integrated photonics networks}
\label{sec:theory}

As mentioned in the introduction, one of the primary advantages of implementing CQFC networks in integrated photonics is the scalability of this platform. Since the SLH formalism, and accompanying automation tools such as the QHDL language \cite{Tezak:2012uc} and the QNET simulation package~\cite{QNET}, implement the analogue of lumped element analysis for quantum optical networks, they are most powerful for analyzing large-scale modular networks that are difficult to simulate from first principles. Practical realizations of such networks require integrated platforms such as superconducting circuits or silicon photonics. However, originally the SLH formalism was developed primarily with bulk-optics implementations in mind, and needs to be reassessed and possibly modified before application to integrated platforms.  

The primary challenges in porting the SLH formalism to integrated photonics stem from the need to capture the range of optical phenomena resulting from electromagnetic field propagation in a nonlinear, dispersive medium. In terms of the assumptions listed at the end of Sec.~\ref{sec:slh}, assumption 3 is the one that needs further examination, since the propagation medium is no longer vacuum. Specific effects that need to be taken into account are dictated by the material substrate and the wavelength of light used in the nanophotonics platform implementation. The dominant physical phenomena present in silicon and silicon nitride integrated photonics at 1550~nm, and absent in bulk-optics networks, were discussed in Sec.~\ref{sec:onchip}. To reiterate, these are: (i) dispersion, (ii) scattering by the medium, including surface roughness scattering, Raman scattering, and Brillouin scattering, and (iii) two-photon absorption and subsequent free carrier generation and heating in the medium. 

In the following, we examine each of the physical effects identified above and assess their impact on the SLH formalism.

\subsection{Dispersion}

In integrated photonics waveguides, chromatic dispersion is a combination of waveguide dispersion and material dispersion. The former is present if the waveguide's guiding properties depend on the light wavelength, and the latter arises from dependence of the material's refractive index on the wavelength. Both types of dispersion can be minimized by engineering the waveguide properties~\cite{Turner:2006gy, Zhang:10, Zhang:2012hz}, however, this engineering is typically very challenging and nontrivial. Therefore, we must examine the consequences of chromatic dispersion on SLH models of integrated photonics devices.

Dispersion needs to be taken into account both in resonant structures and bus waveguides. In the former, it is largely an experimental design issue since it complicates phase matching, which subsequently makes the design of nonlinear elements such as OPOs difficult \cite{Dutt:2013wr}. Resonant structures must be engineered to have required phase matching properties and also be resonant for frequencies of the modes participating in the desired four-wave mixing process. For bus waveguides, dispersion manifests itself as the dependence of the propagation velocity along the bus on the wavelength. 
This is fundamentally incompatible with the assumptions of CQFC and the SLH formalism because strong dispersion can violate the Markov approximation that is necessary for the validity of SLH models (assumption 2 in Section \ref{sec:slh}) \cite{Sta.Bar.etal-2004,CombesKerckhoffSarovar2016}. 
Although there has been some work on mimicking restricted types of dispersive propagation within the standard input-output theory (and SLH) framework \cite{Sta.Wis-2006}, to date there is no general extension to the SLH formalism that can accommodate arbitrary dispersive propagation.

\subsection{Scattering}

Surface roughness scattering leads to conversion of photons from modes of interest into other modes. This can be phenomenologically modeled as a linear loss mechanism that can be easily incorporated into the SLH description of CQFC networks. Specifically, loss in a bus waveguide can be modeled by a fictitious directional coupler (analogous to a beam splitter in bulk-optics) and loss in a localized component can be modeled by an additional fictitious port with vacuum input.

Lack of unidirectional propagation due to spurious impurity-driven back scattering is also a concern in integrated waveguides. If this type of backscattering can be identified, it is possible to model within extensions of the SLH formalism \cite{CombesKerckhoffSarovar2016}, however, this requires precise characterization of backscattering amplitudes in the waveguide.

Effects of nonlinear scattering phenomena such as Raman and Brillouin scattering cannot be modeled as simply as those of surface roughness scattering because the loss coefficient is dependent on the light intensity in these cases. As argued above, these nonlinear scattering effects can be safely ignored in integrated bus waveguides because the field intensities for CQFC applications are typically too small for these effects to be significant. However in resonant structures, especially those constructed to specifically behave in a nonlinear fashion (e.g., integrated-optics implementations of an OPO), nonlinear scattering effects must be taken into account. Since the underlying scattering mechanisms ultimately arise from interactions with crystal phonons, they can be modeled fully quantum mechanically \cite[Secs. 6.4.1, 11.6]{Drummond:2014tr}. As these models show, such scattering produces incoherent loss or gain of population in the modes of interest, as well as phase decoherence. Most significantly for the CQFC framework, only in some special situations can these phenomena be modeled by a coupling to a Markovian reservoir \cite{Drummond:2014tr}, which means that in most cases the effects of these nonlinear scattering processes cannot be modeled within the standard SLH formalism. Accurately incorporating these nonlinear scattering processes within the SLH formalism is an avenue for future work. 

\subsection{Two-photon absorption}

At 1550~nm, TPA is an important process in silicon, but is of less concern in silicon nitride where the band gap is larger. Again, as a nonlinear process it is of concern to us in resonant structures and not in bus waveguides. In resonant structures TPA is a source of nonlinear (intensity dependent) loss, but unlike the case of Raman/Brillouin scattering, this process is Markovian and thus can be described by a Markovian master equation \cite{Gilles:1993bp}. Therefore, we can model the effect of TPA in a localized component by an additional port with vacuum input and coupling quadratic in the amplitude of the component's internal field mode -- \ie the element of the $L$ matrix corresponding to the additional port is proportional to $a^2$, where $a$ is the annihilation operator of the internal field mode.

However, TPA has secondary effects that are not captured by this model. Specifically, TPA typically results in the creation of free carriers, whose concentration affects the refractive index of the material, which in turn changes its nonlinear and guiding properties (and causes dispersion if this change in refractive index is wavelength dependent). Consequently, TPA can dynamically change the underlying parameters of a localized component's system, something that is not captured by the SLH formalism, which usually assumes static parameters. For example, consider an on-chip ring resonator (cavity) that is characterized by the resonance frequency of its fundamental mode and the strength of its coupling to a bus waveguide. If the $Q$ factor of the resonator and input light power are high enough, TPA-induced free carriers in the ring material can shift the resonance frequency. Modifications to the SLH formalism to capture these effects will be essential for accurately modeling resonant integrated optics elements. Some noteworthy progress has recently been made in this direction through the formulation of a quantum model for free carrier dispersion in nanophotonic cavities \cite{Hamerly:2015cb}. Unfortunately although this model is in-principle compatible with the SLH formalism, it is not practical to implement directly using the formalism due to the large number of degrees of freedom that must be accounted for. Hamerly and Mabuchi adopt a semi-classical approach to efficiently simulate their model and it is possible that such methods could be integrated with the SLH formalism to treat such physics.

\section{Experiment: On-chip implementation of CQFC network of two coupled cavities}
\label{sec:device}

\begin{figure}[t!]
\includegraphics[scale=0.7]{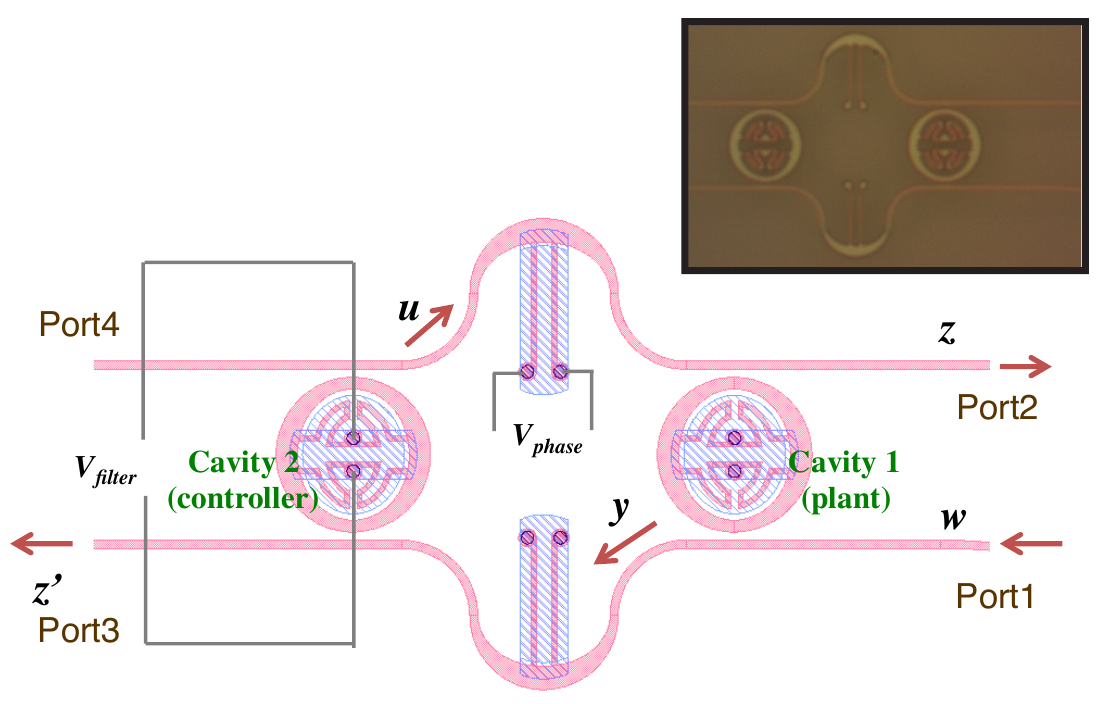}
\caption{\label{fig:coupled_cavities} The coupled cavity device (CCD) consists of two ring resonators coupled by two waveguides. Each of the rings is 6 $\mu$m in diameter and the center-to-center distance between the rings is 15 $\mu$m. A coherent input drive $w$ is applied at one port, and the output field $z$ is monitored at another port. The two remaining ports have a vacuum input and an unmonitored output field $z'$. The SLH model of the CCD is described in Appendix~\ref{sec:appendix}. The inset shows a magnified image of the device.}
\end{figure}

In order to study the differences and similarities between bulk-optics and silicon nanophotonics networks, we fabricated one of the simplest CQFC networks: two cavities coupled in a feedback loop; see Fig.~\ref{fig:coupled_cavities}. We refer to this network as the \emph{coupled cavity device} (CCD), and its implementation was inspired by the disturbance rejection network implemented in bulk optics by Mabuchi in Ref.~\cite{Mab-2008}, and previously theoretically analyzed in Ref.~\cite{James:2008uk}.
 
Figure \ref{fig:coupled_cavities} shows the schematic of the CCD, which consists of two thermally controlled ring resonators (denoted as cavity 1 and cavity 2) coupled by Si nanowire bus waveguides with an integrated thermo-optic phase shifter on each bus waveguide. The device has four actual ports (additional fictitious ports can be used to model losses), two at each cavity. A coherent input drive ($w$) from a widely tunable telecom laser is coupled onto chip from a lensed fiber into port 1. The intensity of the output signal ($z$) at port 2 is monitored off chip using a power meter. The ports of cavity 2 are unmonitored and have vacuum input. The signal in the output field $z$ is a result of interference between the outputs of both cavities since they are coupled. In Refs.~\cite{James:2008uk, Mab-2008}, one of the cavities is treated as a plant system and the other as a controller, in which case the signal propagating from cavity 2 to cavity 1 (field $u$) is viewed as a feedback signal from controller to plant. 

The CCD is controlled using two thermo-optic phase shifters activated by externally applied voltages. One phase shifter, controlled by the voltage $V_{\mathrm{filter}}$, is used to tune the resonance frequency $\omega_c$ of cavity 2, and the other, controlled by the voltage $V_{\mathrm{phase}}$, is used to induce a phase shift $\phi$ in the feedback signal (field $u$). The variable parameters $\omega_c$ and $\phi$ are the primary \emph{in situ} controllable degrees of freedom of the CQFC network implemented by the CCD. During the experiment the probe laser wavelength is swept and the upper phase shifter voltage, $V_{\rm phase}$  is varied from 0 to 18 V for fixed controller $V_{\rm filter}$ bias. We estimate that the resonators have quality factors in the range $1\times 10^3 - 4 \times 10^3$.

The coupled cavities behave in a similar fashion to electromagnetically induced transparency (EIT), where two nearly degenerate resonances can interfere creating sharp null in the transmission spectrum.  $V_{\rm phase}$ acts as a variable coupling  between the two resonant cavities and can be used to shift in wavelength  the interference null in the transmission spectrum (c.f. Refs. \cite{Peng:2014gb, Xu:2006ej}). However, for disturbance rejection we require that the transmission is suppressed at all wavelengths. This can be achieved in the CCD if the linewidth of the controller cavity and the phase shift induced in the feedback signal $u$ can be controlled independently, while the other parameters are fixed \cite{Mab-2008}. Disturbance rejection means that, with suitable parameter values, the output field $z$ is in the vacuum state regardless of the amplitude and phase of the input field $w$. Physically, this results from all the input power being routed to the output port $z'$ due to the interference between the cavities. We are unable to tune the cavity linewidths \emph{in situ} with this generation of the CCD, however, we will evaluate whether disturbance rejection can be achieved with the parameter values determined at fabrication and the \emph{in situ} tuning capabilities we do have. 

\begin{figure*}[t!]
\includegraphics[scale=1.35]{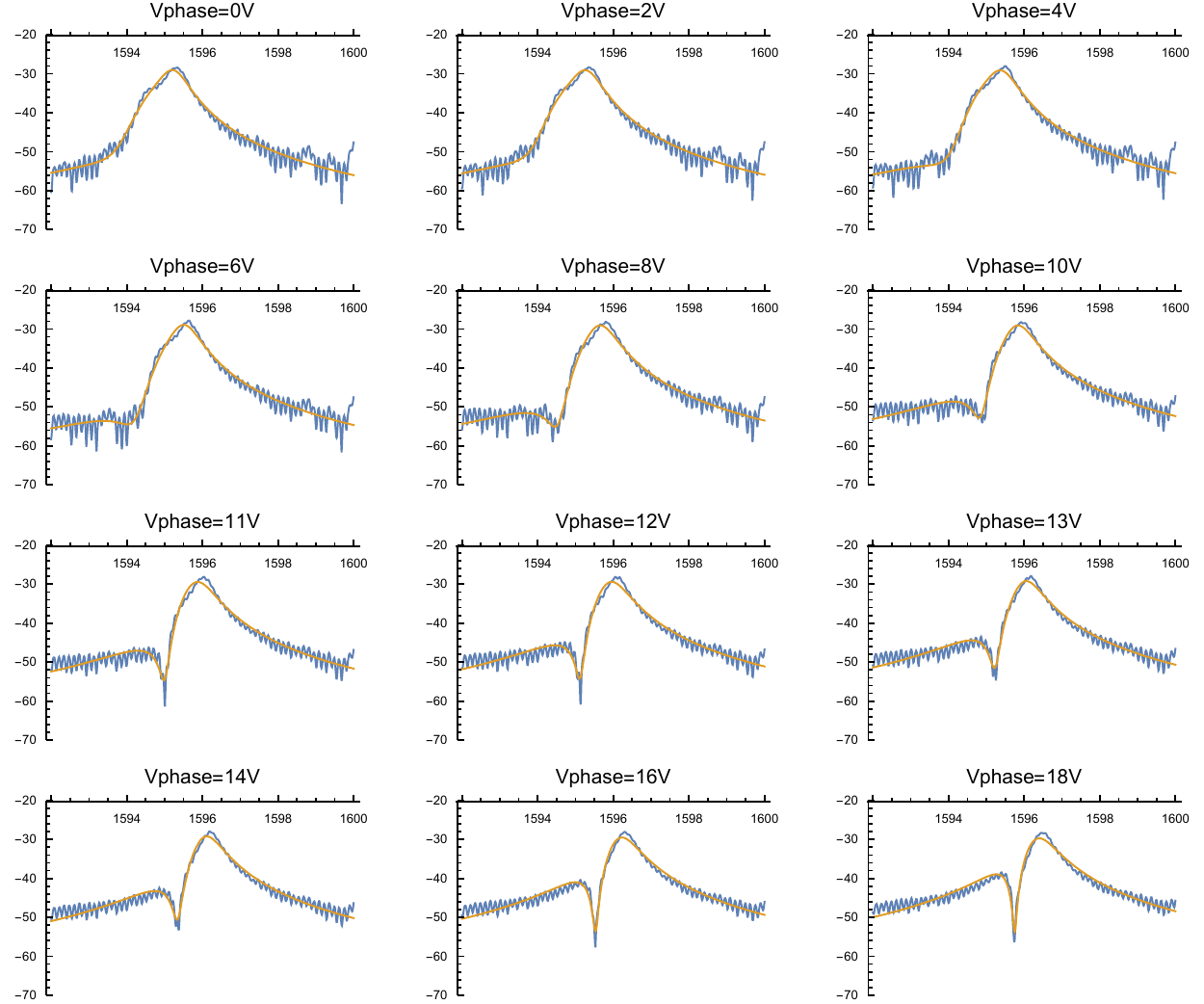}
\caption{\label{fig:output_plots} The spectra of the output field $z$ of the CCD driven by a coherent input field $w$. Each subplot shows the output power spectrum (in dB) for a different value of the voltage $V_{\mathrm{phase}}$ that controls the phase shift $\phi$ induced in the feedback field $u$. The blue lines are data measured in the experiment and the orange lines are theoretical predictions based on the SLH model for the CCD with optimally fitted parameter values. The parameter fitting was done independently for each value of $V_{\mathrm{phase}}$.}
\end{figure*}

In Appendix~\ref{sec:appendix}, we develop a model of the CCD using the SLH formalism. Here, we explicitly list the parameters entering the model:
\begin{enumerate}
\item $\omega_p$: the resonance frequency of the fundamental mode of the plant cavity.
\item $\omega_c$: the resonance frequency of the fundamental mode of the controller cavity.
\item $\kappa$: the coupling rate of both cavities to the bus waveguides. This rate is assumed to be the same for both interfaces of both cavities since it results from an evanescent coupling that is dictated by the distance between the ring resonator and the bus waveguide. This distance is the same, up to fabrication precision, for all interfaces in the CCD.
\item $\gamma_p$: the loss rate of the plant cavity.
\item $\gamma_c$: the loss rate of the controller cavity.
\item $\phi$: the phase shift induced in the feedback signal $u$. 
\item $\eta$: the power loss in the waveguide in which the feedback signal $u$ propagates. This parameter accounts for potential losses due to active control of this waveguide.
\end{enumerate}

It should be noted that if the CCD is driven by a low-power laser (i.e., the input field $w$ is prepared in a low-intensity coherent state) and the cavities remain in the linear regime (i.e., their $Q$ factors are not too large), then the entire network is linear and thus can be modeled by an equivalent classical transfer function as shown in Refs.~\cite{James:2008uk, Mab-2008}. In this case, the input-output relationships predicted by the SLH model and the transfer function agree. The experiment is conducted in this linear regime and therefore the SLH model does not incorporate nonlinearities. In the following, we assess the agreement between this model and experimental data.

Figure~\ref{fig:output_plots} shows the spectra of the output field $z$ when the input CW field $w$ is prepared in a coherent state. The input power (after accounting for on-chip coupling loss) is calibrated to be $\sim 100~ \mu$W, and $V_{\rm filter}$ is held fixed at $1.4$ V. 
Each subplot shows the output spectrum for a different value of the voltage $V_{\mathrm{phase}}$, which controls the phase shift $\phi$ induced on the feedback field $u$. In each subplot, the experimental spectrum is shown together with the spectrum predicted by the theoretical SLH model (developed in Appendix~\ref{sec:appendix}) with parameter values selected to achieve the best fit with the experimental data. The parameter fitting was performed independently for each value of $V_{\mathrm{phase}}$, using a simulated annealing algorithm initialized with a good estimate of the parameter values from knowledge of the chip fabrication details. The fits are seen to be reasonably good, which gives us confidence that the linear model is an accurate representation of the system. 

\begin{figure*}[]
\includegraphics[scale=0.53]{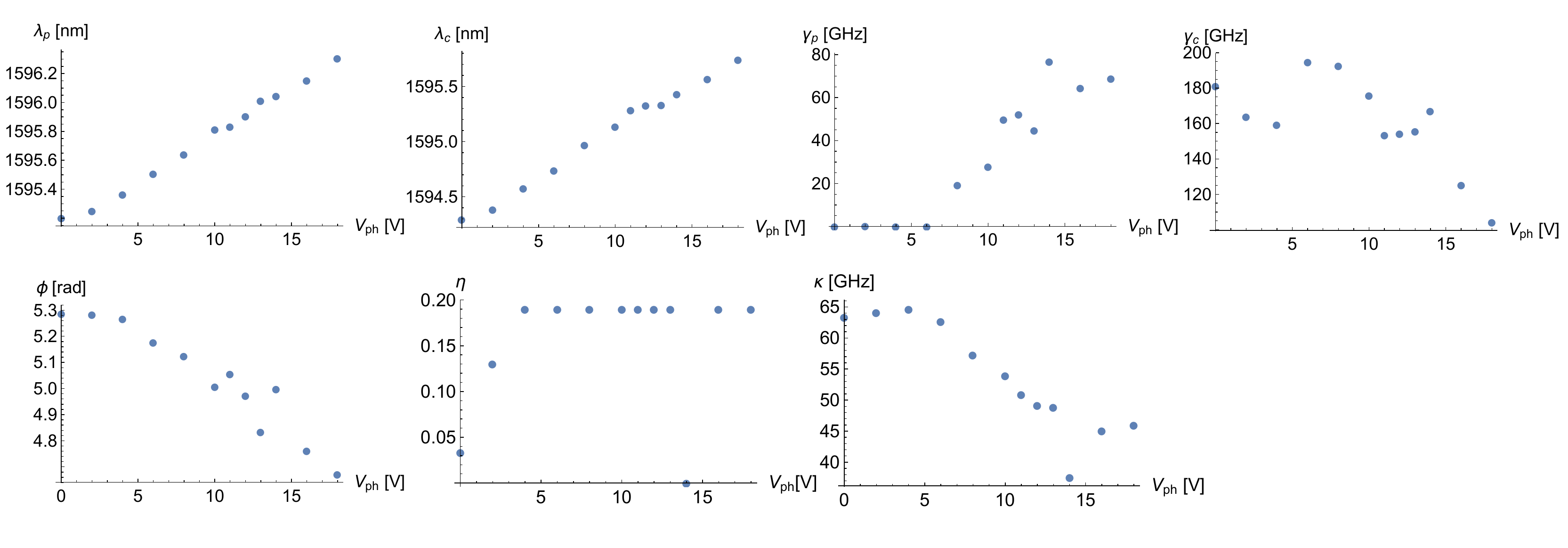}
\caption{\label{fig:param_plots} Parameter values that produce the best fit of the theoretical SLH model to the experimentally measured output spectrum (shown in Fig.~\ref{fig:output_plots}), for each value of the voltage $V_{\mathrm{phase}}$ applied to the phase shifter. Each subplot shows the variation of one of the parameters ($\lambda_p$, $\lambda_c$, $\gamma_p$, $\gamma_c$, $\phi$, $\eta$, and $\kappa$) as $V_{\mathrm{phase}}$ changes between $0$ and $18$~V.}
\end{figure*}

Figure~\ref{fig:param_plots} shows the fitted values of the model parameters ($\lambda_p$, $\lambda_c$, $\gamma_p$, $\gamma_c$, $\phi$, $\eta$, and $\kappa$) obtained for each value of $V_{\mathrm{phase}}$. Note that $\lambda_{c/p} = \frac{2\pi c}{n_{\rm eff} \omega_{c/p}}$ with $n_{\rm eff}=2.85$ being the effective index of the resonator at $1550$ nm. Surprisingly, we see that not only the feedback phase $\phi$, but almost all other parameters change with $V_{\mathrm{phase}}$. This is hardly ideal as it means that the parameters in the system are not independently tunable. In particular, we cannot tune the system to operate in the parameter regime required for disturbance rejection. Physically, the effect of the voltage applied to a phase shifter (where Joule heating changes the material refractive index via the thermo-optic effect) does not seem to be localized to the bus waveguide connecting the cavities; the produced heat also affects the properties of adjacent optical elements, including cavity resonances, cavity-waveguide coupling and cavity losses. This cross-talk effect associated with the physical nature of controls and size characteristics of the nanophotonics platform, is an important issue that distinguishes on-chip implementations of optical networks from their bulk-optics analogues.

We comment on the behavior of each of the parameters in Fig.~\ref{fig:param_plots}. For the most part, the phase shift $\phi$ changes predictably and monotonically with $V_{\mathrm{phase}}$. The power loss in the waveguide, $\eta$, also behaves rather regularly: it increases with the applied voltage for $V_{\mathrm{phase}} \lesssim 5$~V and then stays approximately constant at higher voltages; this behavior implies that the waveguide becomes lossier when the thermo-optic phase shifter is active, but losses saturate at $V_{\mathrm{phase}} \gtrsim 5$~V. On the other hand, the rate of cavity-waveguide couplings, $\kappa$, does not change much for $V_{\mathrm{phase}} \lesssim 5$~V, but decreases when the voltage increases for $V_{\mathrm{phase}} \gtrsim 5$~V; this behavior might indicate that cavity-waveguide interfaces become affected when a higher voltage generates a larger amount of heat. Judging by the changes of cavity resonance wavelengths $\lambda_p$ and $\lambda_c$, the cavity resonances are linearly red-shifted with increasing $V_{\mathrm{phase}}$, and by roughly the same amount for both cavities. The cavity loss rates behave surprisingly dissimilar for the plant and controller resonators. The plant cavity's loss rate $\gamma_p$ is constant for $V_{\mathrm{phase}} \lesssim 6$~V and increases with $V_{\mathrm{phase}}$ for higher voltages. In contrast, the controller cavity's loss rate $\gamma_c$ mostly decreases when the applied voltage increases, but its behavior is significantly non-monotonic and least regular of all the parameters. We were unable to find a high quality, monotonic fit to this parameter even after randomizing the initial state for the simulated annealing optimization algorithm.  

We also performed experiments where both $V_{\rm phase}$ and the resonance frequency of the controller cavity were tuned (via $V_{\rm filter}$), but were also unable to achieve disturbance rejection in this case. We attribute this to the inability to tune the CCD to the disturbance rejection parameter regime due to (i) the cross-talk effect mentioned above inhibiting independent tuning of device parameters, and (ii) the large mismatch between the quality factors of the two cavities due to the much larger linewidth of the controller cavity compared to the plant cavity. 

\section{Discussion}
\label{sec:disc}

In this work, we analyzed the suitability of integrated silicon photonics to serve as a platform for implementing scalable CQFC networks. In addition to summarizing the strengths of silicon photonics for this application, we also outlined the principal challenges to both the theoretical framework and practical implementations. In particular, Sec.~\ref{sec:theory} presented the 
features of the integrated photonics platform that are not yet taken into account by the SLH formalism, which was largely developed with bulk-optics implementations in mind. We identified a number of extensions to the SLH formalism, which are needed to make it more applicable to linear as well as nonlinear integrated optics networks.

In Sec.~\ref{sec:device}, we presented the results of a preliminary experiment that explored an on-chip implementation of a simple CQFC network of two coupled cavities, and analyzed its properties using the SLH formalism. The main lesson that we learned from this analysis is that \emph{in situ} controls applied to one component of the nanophotonic device significantly affected properties of the rest of the components. This cross-talk is a result of the small size of the device and the physical nature of the controls that utilize Joule heating to manipulate optical properties. Therefore, it is important to understand all the impacts of \emph{in situ} controls including thermal effects due to the heat transfer and effects on carrier concentration, both of which can change properties of integrated network components. It is also clear that it is advantageous to use \emph{in situ} control mechanisms that act locally, and have as few side effects, as possible.

The fast progress in the field of nanophotonics technology is likely to generate new advances that will be beneficial to integrated optics implementations of CQFC networks. In particular, recent advances in non-silicon CMOS-compatible platforms utilizing low-loss materials such as silicon nitride and Hydex will likely enable more versatile integrated components, especially for nonlinear optics~\cite{Moss:2013kv}. For example, the Hydex platform was used to implement an integrated photon pair source~\cite{Reimer:2014} employing an above-threshold OPO, a key nonlinear optics element. Another promising development is the use of hydrogenated amorphous silicon (a-Si:H) whose nonlinear optical properties in the telecommunications band (including ultrahigh optical nonlinearity, low nonlinear loss, and reduced impact of free-carrier processes) are superior to those of undoped crystalline silicon~\cite{Kuyken:2011nl}. The a-Si:H platform was used to demonstrate integrated photonics implementations of various nonlinear optical processes, including parametric amplification~\cite{Kuyken:2011pa}, four-wave mixing for low-power optical frequency conversion~\cite{Wang:2012a, Wang:2012b, Matres:2013}, and cross-phase modulation for all-optical switching~\cite{Pelc:2014}. The incorporation of new materials into CMOS compatible processes and heterogenous integration into the Si photonics platform will greatly expand the integrated quantum optics toolbox, and enable the construction increasingly complex quantum optical networks. 

\acknowledgments
MS would like to acknowledge helpful conversations with Joseph Kerckhoff and Joshua Combes.
This work was supported by the Laboratory Directed Research and Development program at Sandia National Laboratories. Sandia is a multi-program laboratory managed and operated by Sandia Corporation, a wholly owned subsidiary of Lockheed Martin Corporation, for the United States Department of Energy's National Nuclear Security Administration under contract DE-AC04-94AL85000. 

\pagebreak
\newpage

\begin{widetext}
\appendix

\section{SLH model for the coupled cavity device}
\label{sec:appendix}

In order to develop the theoretical model of the CCD depicted in Fig.~\ref{fig:coupled_cavities}, we use a schematic representation of the equivalent network shown in Fig.~\ref{fig:coupled_cavities_schematic}. Note that the network shown in Fig.~\ref{fig:coupled_cavities_schematic} includes additional fictitious components (a beam splitter and two mirrors) that are used to model losses. This network can be decomposed as a concatenation product \cite{Gou.Jam-2009, Gough:2012fl} of several components in series. Specifically, the SLH triple for the CCD, $G_{\mathrm{CCD}}$, is given by:
\bqa
G_{\mathrm{CCD}} = \big[ (G_{p2}\boxplus G_1) \lhd G_{\eta} \lhd (G_\phi \boxplus G_1) \lhd (G_{c2}\boxplus G_1) \big] \boxplus \big[ G_{c1} \lhd G_{p1} \lhd G_w \big] \boxplus G_{p3} \boxplus G_{c3},
\label{eq:slh_breakdown}
\eqa
where the components are specified by the following SLH triples:
\begin{align*}
& G_{p1} = \left( 1, \sqrt{\kappa} a, \omega_p a\dg a \right), \\
& G_{p2} = \left( 1, \sqrt{\kappa} a, 0 \right), \\
& G_{p3} = \left( 1, \sqrt{\gamma_p} a, 0 \right), \\
& G_{c1} = \left( 1, \sqrt{\kappa} b, \omega_c b\dg b \right). \\
& G_{c2} = \left( 1, \sqrt{\kappa} b, 0 \right), \\
& G_{c3} = \left( 1, \sqrt{\gamma_c} b, 0 \right), \\
& G_w = \left( 1, \alpha, 0 \right), \\
& G_{\phi} = \left( e^{i\phi}, 0, 0 \right), \\
& G_1 = \left( 1, 0, 0 \right), \\
& G_{\eta} = \left( \left[
\begin{array}{cc}\sqrt{1-\eta} & \sqrt{\eta} \\ \sqrt{\eta} & \sqrt{1-\eta}\end{array} \right],0,0\right).
\end{align*}
Here, $a$ is the annihilation operator for the fundamental mode of the plant cavity, of frequency $\omega_p$, and $b$ is the annihilation operator for the fundamental mode of the controller cavity, of frequency $\omega_c$. $G_{p1}$ and $G_{p2}$ represent the ``mirrors'' of the plant cavity that couple to the bus waveguides. Similarly, $G_{c1}$ and $G_{c2}$ represent the ``mirrors'' of the controller cavity that couple to the bus waveguides. We assume that all these mirrors have the same leakage rate $\kappa$ (the leakage rate $\kappa$ of a cavity mirror relates to its power transmittance $T$ as $\kappa = c T/(2 n_{\mathrm{eff}} \ell)$, where $c$ is the speed of light, $n_{\mathrm{eff}}$ is the refractive index of the cavity medium, and $\ell$ is the cavity length). $G_{p3}$ and $G_{c3}$ represent the fictitious ``mirrors'' that leak light and are used to model intrinsic losses in the plant cavity and the controller cavity, respectively; these ``mirrors'' have leakage rates $\gamma_p$ and $\gamma_c$, respectively. $G_w$ represents the drive field applied to the plant cavity (signal $w$), which is assumed to be prepared in a coherent state with complex amplitude $\alpha$. $G_{\phi}$ represents the phase shifter that induces the phase shift $\phi$ in the feedback field $u$, and $G_1$ represents a simple passthrough. Finally, $G_{\eta}$ represents the fictitious beam-splitter that models the loss in the waveguide linking the controller and plant. The power loss in the waveguide is $\eta$, so the $\sqrt{1-\eta}$ portion of the field amplitude is transmitted. We see from this component breakdown that there are seven free parameters in this model: $\omega_p$, $\omega_c$, $\kappa$, $\gamma_p$, $\gamma_c$, $\eta$, and $\phi$.

Equation~(\ref{eq:slh_breakdown}) expresses the power of the SLH formalism; it captures the modular nature of the optical network and also gives us a prescriptive formula for how to model the properties of the entire device by knowing properties of each module. For simple networks like the one considered here, it is possible to model each component and deduce the appropriate product by examination. However, there exist systematic and prescriptive techniques to achieve the same SLH component breakdown for an arbitrary network~\cite{Gou.Jam-2009, Gough:2012fl}. Furthermore, the QNET software package~\cite{QNET} enables complete automation of this task, starting from a specification of the network in terms of its physical components.

\begin{figure}[b]
\includegraphics[width=0.6\linewidth]{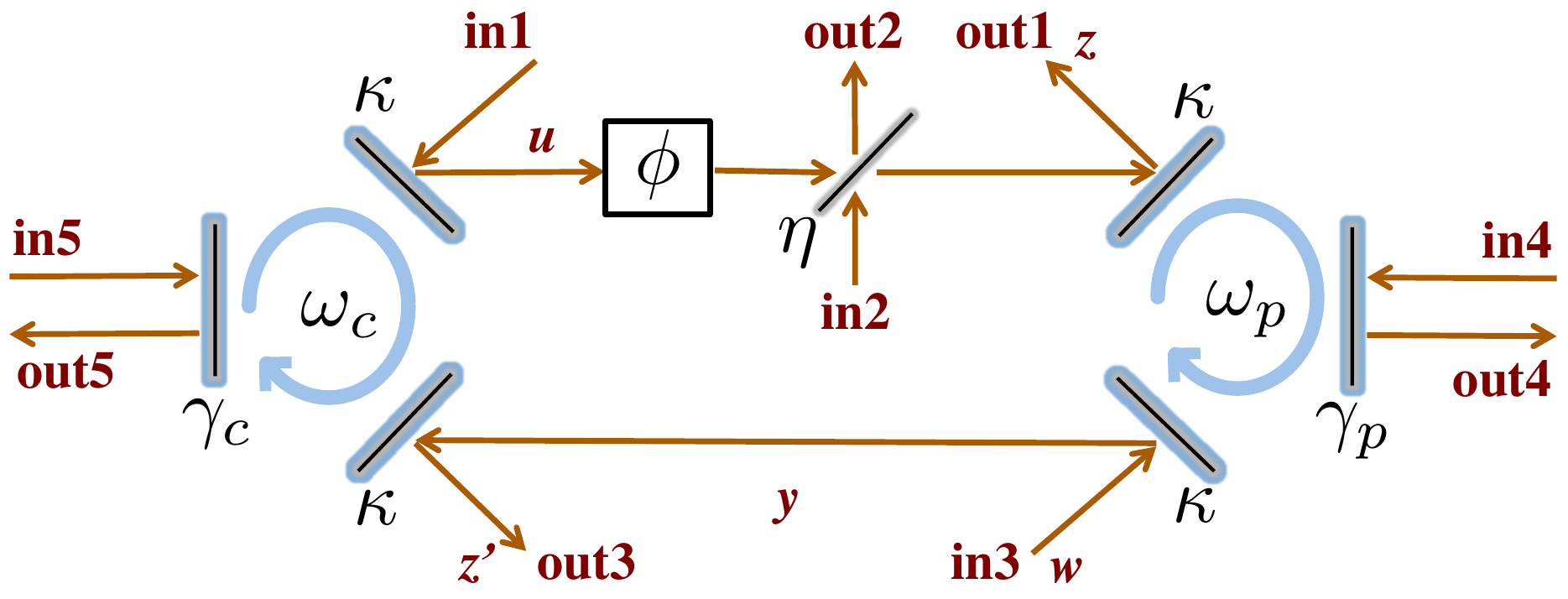}
\caption{\label{fig:coupled_cavities_schematic} A schematic representation of the CCD, used to generate a lumped element model. Each ring resonator is modeled as a triangular cavity with mirrors that couple to two waveguides with leakage rate $\kappa$, and one fictitious mirror that represents intrinsic loss with leakage rate $\gamma_{p/c}$. The cavity on the right is the plant and the one of the left is the controller. The driving signal $w$ and the monitored signal $z$ are indicated. The feedback signal $u$ undergoes a phase shift $\phi$ and power loss $\eta$. The overall network has five input ports and five output ports, all of which are indicated; the driving field $w$ enters at input port 3 and the monitored field $z$ exits at output port 1; all input ports except 3 have vacuum fields. Note that there is a re-labeling of the ports with respect to Fig. 1 in the main text.}
\end{figure}

Carrying out the series and concatenation products in \erf{eq:slh_breakdown}, yields the SLH triple representing the entire CCD:
\bqa
G_{\mathrm{CCD}} &=& \left( \left[\begin{array}{ccccc}\sqrt{1-\eta}e^{i\phi} & \sqrt{\eta} & 0 & 0 & 0\\ \sqrt{\eta}e^{i\phi} & \sqrt{1-\eta} & 0 & 0 & 0\\ 0 & 0 & 1 & 0 & 0 \\ 0 & 0 & 0 & 1 & 0 \\ 0 & 0 & 0 & 0 & 1\end{array}\right],  \left[\begin{array}{c} \sqrt{\kappa}( a + \sqrt{1-\eta}e^{i\phi} b) \\ \sqrt{\kappa\eta}e^{i\phi}b \\ \sqrt{\kappa}(a+b) + \alpha \\ \sqrt{\gamma_p}a \\ \sqrt{\gamma_c}b \end{array}\right], \right. \nn \\
 && \left. \omega_p a\dg a + \omega_c b\dg b + \frac{\sqrt{\kappa}}{2i}\left(a\dg \alpha - a \alpha^*\right) + \frac{\kappa}{2i}\big[(1-e^{-i\phi}\sqrt{1-\eta})a\dg b - (1-e^{i\phi}\sqrt{1-\eta})ab\dg\big] \right)
\label{eq:G_ccd}
\eqa
The ordering of elements in matrices $S$ and $L$ corresponds to the numbering of input and output ports in Fig.~\ref{fig:coupled_cavities_schematic}, i.e., the matrix element $S_{i j}$ is the scattering amplitude from input port $i$ to output port $j$, and the matrix element $L_i$ represents the field at output port $i$.

Given the SLH triple (\ref{eq:G_ccd}) for the CCD, one can explicitly see that (i) the Hamiltonian includes a coupling term representing an effective interaction between the plant cavity and the controller cavity, induced by the feedback, and (ii) the signal field $z$ at output port~1 is produced by an interference (i.e., composed of a linear combination) of the two cavity fields: $a + \sqrt{1-\eta} e^{i\phi} b$. The observation made in Refs.~\cite{James:2008uk} and \cite{Mab-2008} is that by appropriate choice of parameters, especially the phase shift $\phi$, this output can be made zero regardless of the driving field $w$ at input port~3. 

We can explicitly write output field at port 1 (out1 or $z$) as
\bqa
dB_{\rm out1} &=& L_1(t) dt + \sum_l S_{1l} dB^{\rm in}_{l} \nn \\
&=& \sqrt{\kappa}(a + \sqrt{1-\eta}e^{i\phi} b) dt + \sqrt{1-\eta}e^{i\phi} dB_{\rm in1} + \sqrt{\eta} dB_{\rm in2}
\eqa
It can be verified that the input-output relationship predicted by this SLH model is identical to that predicted by the transfer function derived in Ref.~\cite{Mab-2008}. This is to be expected since the system is linear and the input is in a coherent state.

More generally, for CQFC networks with non-linear optical components, the SLH formalism can be used to model quantum optical phenomena that have no classical analogues. For example, the SLH model of a network of two coupled OPOs~\cite{Gough:2009} predicts interferences that enable to achieve a degree of control over properties of the output field, such as its squeezing spectrum~\cite{Crisafulli:2013dq}, which would be impossible without the coherent feedback.

\end{widetext}

\end{document}